\documentclass[11pt,preprint,nofootinbib]{revtex4}%
\usepackage[T1]{fontenc}
\usepackage{amsmath}
\usepackage{amssymb}
\usepackage{esint}
\usepackage{amsfonts}
\usepackage{graphicx,color}
\usepackage{slashed}
\usepackage{young}
\usepackage[vcentermath]{youngtab}
\usepackage[utf8]{inputenc}
\usepackage{tensor}
\usepackage{etoolbox}
\usepackage{bbm}
\usepackage{soul}
\usepackage{ulem}

\usepackage{epic,eepic,pstricks}
\usepackage{mathrsfs}

\makeatletter

\newcommand{\be}{\begin{equation}}
\newcommand{\ee}{\end{equation}}
\newcommand{\bea}{\begin{eqnarray}}
\newcommand{\eea}{\end{eqnarray}}

\renewcommand{\tilde}{\widetilde}
\renewcommand{\hat}{\widehat}

\renewcommand{\d}{\partial}

\@addtoreset{equation}{section}

\def\cL{\mathcal{L}}

\newcommand*\xbar[1]{%
  \hbox{%
    \vbox{%
      \hrule height 0.5pt % The actual bar
      \kern0.3ex%         % Distance between bar and symbol
      \hbox{%
        \kern-0.0em%      % Shortening on the left side
        \ensuremath{#1}%
        \kern-0.0em%      % Shortening on the right side
      }%
    }%
  }%
} 
\pdfpageheight\paperheight
\pdfpagewidth\paperwidth

\pdfoutput=1 % if your are submitting a pdflatex (i.e. if you have
    \patchcmd{\maketitle}{\@fpheader}{}{}{}

\begin{document}

\title{The asymptotic structure of gravity at spatial infinity in four spacetime dimensions\footnote{Invited contribution to the  special issue of the ``Proceedings of the Steklov Institute of Mathematics'' dedicated
to the 80th anniversary of Andrei Slavnov.} }
\author{Marc Henneaux${}^1$\footnote{On leave of absence from Coll\`ege de France, 11 place Marcelin Berthelot, 75005 Paris, France} and C\'edric Troessaert${}^2$}
\affiliation{${}^1$Universit\'e Libre de Bruxelles and International Solvay Institutes, ULB-Campus Plaine CP231, B-1050 Brussels, Belgium}
%\affiliation[b]{Coll\`ege de France, 11 place Marcelin Berthelot, 75005 Paris, France}
\affiliation{${}^2$Max-Planck-Institut f\"{u}r Gravitationsphysik (Albert-Einstein-Institut),
Am M\"{u}hlenberg 1, \\ DE-14476 Potsdam, Germany}

%\abstract
\begin{abstract}
{A review of our results on the asymptotic structure of gravity at spatial infinity in four spacetime dimensions is given.  Finiteness of the action and integrability of the asymptotic Lorentz boost generators are key criteria that we implement through appropriate boundary conditions.  These conditions are ``twisted parity conditions'', expressing that the leading order of the asymptotic fields obey strict parity conditions under the sphere antipodal map up to an improper gauge transformation.  The asymptotic symmetries are shown to form the infinite-dimensional BMS group, which has a non trivial action. The charges and their algebra are worked out. The presentation aims at being self-contained and at possessing a pedagogical component.}
\end{abstract}

%\pacs{11.30.Ly,11.30.Pb,04.65.+e} CHECK

%\makeatother

%\begin{document}
%\maketitle \flushbottom
\maketitle

\section{Introduction}
\setcounter{equation}{0}

This paper is dedicated to Andrei Alekseevich Slavnov, colleague and friend, on the occasion of his 80th birthday.  His insight on symmetries and his interest in the infrared structure of gauge theories (see e.g. \cite{Slavnov:1981mre}) remain invaluable sources of inspiration.

We review here our recent work on the asymptotic structure of gravity and gauge theories at spatial infinity \cite{Henneaux:2018cst,Henneaux:2018gfi,Henneaux:2018hdj,Henneaux:2018mgn,Henneaux:2019yqq}, in particular \cite{Henneaux:2018hdj}.  Our analysis  investigates the asymptotic dynamics of the fields on spacelike hypersurfaces that approach spacelike (Cauchy) hyperplanes at infinity.  It uses the Hamiltonian formalism of general relativity \cite{Dirac:1958sc,Arnowitt:1962hi} as its main tool and finds its roots in the pioneering paper \cite{Regge:1974zd}.  Our central goal is to understand the emergence at spatial infinity of the BMS symmetry, originally discovered at null infinity \cite{Bondi:1962px,Sachs:1962wk,Sachs:1962zza}.

There are various motivations for studying the asymptotic structure of gravity at spatial infinity, on spacelike hypersurfaces.   (i) First, the existence of null infinity is a rather delicate dynamical question (what are the conditions on Cauchy data that will lead to a null infniity with sufficient smoothness properties?) and it seems therefore to be a good strategy to investigate the symmetries of the theory in a context that is independent from these queries.  This puts the discussion on a firmer footing, and also, it disentagles the existence of the BMS symmetry -- a symmetry of the theory -- from gravitational radiation, known to potentially destroy the usual smoothness requirements made at null infinity \cite{Christodoulou:1993uv}.  (ii) Second, earlier studies of the Hamiltonian structure at spatial infinity did not identify the BMS group as group of physical symmetries, leading to a tension between spatial infinity and null infinity results.  It seems important to eliminate these previous tensions. 
(iii) Finally the quantum formulation of the theory and questions such as Ward identities or dressing of physical states are naturally investigated on Cauchy hypersurfaces.

The question of formulating consistent boundary conditions for given dynamical laws does not have a unique answer.  We have provided in \cite{Henneaux:2018cst} and \cite{Henneaux:2018hdj} two different sets of boundary conditions for gravity at spatial infinity that are invariant under the BMS group and that yield a well-defined and non trivial canonical action of that group.  Both sets are acceptable from the Hamiltonian points of view but we believe that the boundary conditions of \cite{Henneaux:2018hdj} have some advantages: (i) They accomodate the Taub-NUT solution \cite{Misner:1963fr,Bunster:2006rt}.  (ii) They do not lead to logarithmic divergences in the curvature as one goes to null infinity (although subleading polylogarithmic terms will generically be present) \cite{Troessaert:2017jcm}, \cite{Henneaux:2018cst,Henneaux:2018gfi,Henneaux:2018hdj}.

As \cite{Regge:1974zd} showed, the standard fall-off of the spatial metric and its conjugate momentum generally assumed at spatial infinity does not lead to a well-defined canonical formulation of the theory.  The charges of the homogeneous Lorentz group generically diverge, among other difficulties.  One needs therefore to strengthen this fall-off.  The authors of \cite{Regge:1974zd} cured the problem by imposing parity conditions on the leading components of the metric and its momentum: under the antipodal map $x^i \rightarrow - x^i$, the leading metric components are even while the leading components of their momenta are odd (in asymptotically cartesian coordinates). It follows from these parity conditions that both the electric and magnetic components of the Weyl tensor have leading terms that are even under parity \cite{Henneaux:2018hdj}.

Now, the metric components and their momenta are not invariant under diffeomorphisms.  The main idea of \cite{Henneaux:2018hdj} is to impose parity conditions but with a twist.  That is, one does not impose that the leading orders of the metric components (respectively,  their conjugate momenta) be strictly even (respectively, strictly odd), but only that they should be even (respectively odd) up to a diffeomorphism that need not have a definite parity\footnote{This is equivalent to strict parity conditions on the leading orders of the components of the (invariant) Weyl tensor, namely, that they should be even in cartesian coordinates,  if one imposes also extra conditions on some BMS-invariant asymptotic fields characterizing the structure at spatial infinity  - see appendix A.2 of \cite{Henneaux:2018hdj} and Conclusions.}.

If the diffeomorphisms needed to enforce the strict parity conditions were transformations not changing the physical state of the system, one could impose strict parity conditions on the non-invariant metric components and their momenta.  However, as clearly explained in the insighful work \cite{Benguria:1976in}, one must distinguish between ``proper'' and ``improper'' gauge transformations.  While the former have zero charges and correspond to redundancies in the description of the system, the latter have non vanishing charges and are physical transformations that cannot be gauge fixed.  For that reason, they must be retained in the asymptotic form of the fields and it would be incorrect to set them to zero.

This conceptually simple and perhaps even obvious extension of the work of  \cite{Regge:1974zd} (``parity conditions involving a twist given by improper diffeomorphisms'') completely reconciles the symmetry analyses at spatial and null infinity.  One can prove in particular that it is the same BMS group that acts both at spatial infinity and at null infinity, expressed in different parametrizations that can be explicitly related \cite{Troessaert:2017jcm}, \cite{Henneaux:2018hdj}.   The derivation of the Hamiltonian formalism with these more general parity conditions is not entirely straightforward on the technical side, however, because the Hamiltonian variables are not diffeomorphism invariant. This development was carried out in \cite{Henneaux:2018hdj}. 

 It is the purpose of this article to systematically review the results of \cite{Henneaux:2018hdj} in a self-contained manner, with a somewhat pedagogical emphasis that puts new light on some features of the derivation.

\section{Action and notations}
Since we will use the Hamiltonian formulation of general relativity throughout, we briefly recall its key features. The hamiltonian action of pure gravity in $4$ spacetime dimensions can be written as
\begin{gather}
	S[g_{ij}, \pi^{ij}, N, N^i]  =\int dt \left\{ \int d^3x \left( \pi^{ij} \d_t g_{ij}  - N^i \mathcal
		H^{grav}_i - N \mathcal H^{grav} \right) - B_{\infty}  \right\},\\
	\mathcal H^{grav}  = - \sqrt g R + \frac{1}{\sqrt g} (\pi^{ij} \pi_{ij} -
	\frac{1}{2} \pi^2),\quad
	\mathcal H^{grav}_i  = -2 \nabla_j \pi^j_i.
\end{gather}
where $g_{ij}$ is the $3$-dimensional metric on the (spacelike) constant  time  slices,   $g$ its determinant, $\pi^{ij}$ its conjugate momentum, $N$ the lapse and $N^i$ the shift \cite{Dirac:1958sc,Arnowitt:1962hi}.  The spatial covariant derivative with respect to the metric $g_{ij}$ is denoted by $\nabla_i $.  The scalar curvature of $g_{ij}$ is $R$. The term $B_{\infty}$ is a boundary term on the $2$-sphere at spatial infinity that depends on the asymptotic values of the lapse and of the shift (see Eq (\ref{eq:BounTermAction0}) below). Varying the action with respect to the lapse and the shift yields the ``Hamiltonian'' and ``momentum'' constraints $\mathcal H^{grav} \approx 0$, $\mathcal H^{grav}_i  \approx 0$. The lapse and the shift act therefore as Lagrange multipliers for the constraints.

We are interested in spacetimes that approach the Minkoswski spacetime at large distances.  The precise rates of decay of the various fields will be given in the next section.  Coordinates $x^i$ ($i = 1, 2, 3$)  will refer from now on to asymptotically cartesian coordinates (except in Appendix \ref{app:2+1}). Asymptotically polar coordinates will be denoted by $(r, x^A)$, where $r^2 = \sum_i \left(x^i \right)^2$ and where the $x^A$ ($A = 1,2$) are coordinates on the unit sphere (referred to as the angles). The flat metric to which the spatial metric tends at infinity reads
$\delta_{ij} dx^i dx^j = dr^2 + r^2 \xbar \gamma_{AB} dx^A dx^B$
where $\xbar \gamma_{AB}$ is the round metric on the unit sphere.  Large distances correspond to $r \rightarrow \infty$.  
Functions on the unit sphere are denoted $\xbar f(x^A)$ or $\xbar f(\mathbf{n}^i)$ where $\mathbf{n}^i$ is the unit normal to the spheres. Quantities with an overbar will always be quantities defined on the $2$-sphere, depending  only on the angles (but as it will be clear from context, there will also be quantities on the $2$-sphere for which we shall not use the overbar notation).  So  $\xbar v^A$ is for instance a vector on the $2$-sphere.  The indices $A, B, ...$ on fields with an overbar  will be lowered and raised with the unit metric $\xbar\gamma_{AB}$ and its inverse $\xbar \gamma^{AB}$, introducing no $r$-dependence.  So, $\xbar v_A = \xbar \gamma_{AB} \xbar v^B$, $\xbar v^A = \xbar \gamma^{AB} \xbar v_B$, etc.

Parity is the transformation $x^k \rightarrow -x^k$, which induces the sphere antipodal map $\mathbf{n}^k \rightarrow -\mathbf{n}^k$ together with $r \rightarrow r$.  This transformation has a well defined meaning asymptotically.
In terms of the angles $x^A$, the sphere antipodal map is symbolically written $x^A \rightarrow - x^A$ (it is actually $\theta \rightarrow \pi - \theta$ and $\varphi \rightarrow \varphi + \pi$ in usual $(\theta, \varphi)$ angular variables, where $\xbar \gamma_{AB} dx^A dx^B = \left(d\theta \right)^2 + \sin^2 \theta \left( d \varphi \right)^2$).

\subsubsection*{Flat space geometry}

Since this is relevant to leading order in the $r \rightarrow \infty$ limit, it is useful to recall a few elementary properties of spheres centered at the origin in Euclidean geometry, with induced metric $\mathring \gamma_{AB} = r^2 \xbar \gamma_{AB}$.

The components of the unit normal $\mathbf{n}$ to the spheres are given by $n^i = \frac{ \d x^i}{\d r}$.  We introduce the frame $\{ n^i \equiv  \frac{ \d x^i}{\d r} = \frac{x^i}{r} , e_A^i  \equiv \frac{ \d x^i}{\d x^A} \} $ tangent to the spherical coordinate lines.  The dual frame is $\{ n_i = \frac{\d r}{\d x^i} , {e^A}_i = \frac{\d x^A}{\d x^i}\} $.  One has
$$ \delta_{ij}  n^i n^j = 1, \quad \delta_{ij}  n^i  e_A^j  = 0, \quad \delta_{ij}  e_A^i e_B^j = r^2 \xbar \gamma_{AB}, \quad \delta^{ij}  n_i n_j = 1, \quad \delta^{ij}  n_i {e^A}_j = 0, \quad \delta^{ij}  {e^A}_i  {e^B}_j = \frac{1}{r^2} \xbar \gamma^{AB}.$$
One easily gets
$$ \d_j n^i = \frac{1}{r} \Big( \delta^i_j - n^i n_j \big), \qquad \d_i e^A_{\; k} = - \frac{1}{r} \Big(n_k e^A_{\; i} + n_i e^A_{\; k}\Big) -  ^{(2)} \! \! \xbar \Gamma^A_{\; \; BC} e^B_{\; i} e^C_{\; k} $$
where $^{(2)} \!  \xbar \Gamma^A_{\; \; BC} $ is the Christoffel symbol for the metric $\xbar \gamma_{AB}$.

The covariant derivative with respect to the $3$-dimensional flat metric is denoted by $\mathring \nabla_i$.   Quite generally, the symbol $\mathring {}$ over a quantity refers to Euclidean space.  The $2$-dimensional covariant derivatives with respect to the metrics $\mathring \gamma_{AB}$ and $\xbar \gamma_{AB}$ are respectively denoted by $\mathring D_A$ and $\xbar D_A$\footnote{One has in fact $\mathring D_A=\xbar D_A$ since the two metrics are related by rescalings.  However, in an asymptotically flat  curved geometry, the metric $\gamma_{AB}$ of the constant $r$ spheres is given by $\gamma_{AB} = \mathring \gamma_{AB} + $ ``subleading terms'' so that the covariant derivative $D_A$ with respect to $\gamma_{AB}$   differs from $\xbar D_A$ by subleading terms.  The metric $\xbar \gamma_{AB}$ is still equal to $\lim_{r \rightarrow \infty} \frac{1}{r^2} \gamma_{AB}$, characterizing therefore the leading behaviour of $\gamma_{AB}$.}.  Their respective curvatures $^{(2)} \!  \mathring R$ and $^{(2)} \!  \xbar R$ are
$
^{(2)} \!  \mathring R_{AB} = ^{(2)} \!  \xbar R_{AB} = \xbar \gamma_{AB}$, $^{(2)} \! \mathring R = \frac{2}{r^2}$,  $^{(2)} \!  \xbar R = 2
$.

The extrinsic curvature $\mathring K_{AB}$ of the constant $r$ spheres in Euclidean space is defined by
$
e_A^i \mathring \nabla_i n^j = - \mathring K_A ^{\; \, B} \, e_B^j
$.
One finds $ \mathring K_{AB}  \equiv \mathring \gamma_{BC} \mathring K_A ^{\; \, C}= - r \xbar \gamma _{AB}$, $\mathring K_A ^{\; \, B} = - \frac{1}{r} \delta_A ^{\; \, B}. $
The extrinsic curvature fulfills the Gauss-Codazzi equations
$
\mathring K_{A}^{\; B} \mathring K^{A}_{\; B} - \mathring K^2 +  ^{(2)} \! \! \mathring R = 0$,  $\partial_A \mathring K - \mathring D_B \mathring K^B_{\; A} = 0$.

\section{Boundary conditions} 
We start by taking as asymptotic behaviour of the metric and its conjugate momentum the following fall-off in cartesian coordinates,
\be
	\label{eq:asymptgrav-1}
		g_{ij} = \delta_{ij}  + O(r^{-1}) , \qquad
		\pi^{ij} =O(r^{-2}) ,
\ee
This asymptotic fall-off must be strengthened.  Indeed, two problems arise if one does not impose further conditions on the  leading $O(\frac{1}{r})$-order of the metric and $O(\frac{1}{r^2})$-order of the conjugate momenta.  These are:
\begin{itemize}
\item First, the kinetic term in the action diverges since its leading piece behaves as $\frac{1}{r^3}$, which generically produces a logarithmic divergence upon integration over space (which involves $\int r^2 dr$, yielding $\int \frac{dr}{r}$).  Finiteness of the action is an important physical requirement.  
\item Asymptotic boosts have non-integrable generators, i.e., are not canonical transformations (see Subsection \ref{sec:Integrability}).  Since asymptotic flatness should imply invariance under boosts, this again seems to be an undesirable situation.
\end{itemize}

One way to solve the first difficulty would be to require that the leading terms of the metric and its conjugate momentum have opposite parity properties under the antipodal map, specifically \cite{Regge:1974zd}, 
 $$
h_{ij} \equiv g_{ij} - \delta_{ij} = \frac{\xbar h_{ij}(\mathbf{n}^k)}{r} + O(\frac{1}{r^2}), \quad \xbar h_{ij} (-\mathbf{n}^k) = \xbar h_{ij} (\mathbf{n}^k)
$$
and
$$
\pi^{ij} = \frac{\xbar \pi^{ij}(\mathbf{n}^k)}{r^2} + O(\frac{1}{r^3}), \quad \xbar \pi^{ij} (-\mathbf{n}^k) = - \xbar \pi^{ij} (\mathbf{n}^k).
$$ 
The leading term in $\pi^{ij} \d_t g_{ij}$ is then of the form $f(\mathbf{n}^k) \frac{1}{r^3}$ where $f(\mathbf{n}^k)$ is an odd function on the sphere, and so the logarithmic divergence $\int \frac{dr}{r}$ is multiplied by the integral $\int \sin \theta f(\theta, \varphi) d \theta  d \varphi$, which is zero.

As emphasized in \cite{Henneaux:2018hdj}, these strict parity conditions must be relaxed (without loosing the good property that the kinetic term in the action is finite), because in order to reach a coordinate system where the parity conditions hold, one might have to perform an ``improper gauge transformation'' \cite{Benguria:1976in}, and such gauge fixings are not permissible.

We have suggested in \cite{Henneaux:2018hdj} to replace the strict parity conditions by parity conditions involving a ``twist'' given by an improper gauge transformations\footnote{The improper gauge transformations written in Hamiltonian form are given by (\ref{eq:Transf0gij}) and (\ref{eq:Transf0piij}). The order one function $\zeta$ and vector $\zeta^i$ are the asymptotic normal and tangential components of the improper gauge transformations (these involve no boost or rotation). To leading order, only the $\cL_{\xi^m} g_{ij}$ piece of $\delta g_{ij}$ and the $g^{\frac12} (\zeta^{\vert ij} - g^{ij} \zeta^{\vert m}_{\; \; \; \vert m})$ piece of $\pi^{ij}$ are relevant.}.  More precisely, we allow both an even component $(\xbar h_{ij})^{even}(\mathbf{n}^k)$ and an odd component $(\xbar h_{ij})^{odd}(\mathbf{n}^k)$  in $\xbar h_{ij}(\mathbf{n}^k)$, but while the even component  is arbitrary, the odd component is requested to take the specific form of a diffeomorphism (which goes to a constant at infinity in order to preserve the asymptotic decay in $r^{-1}$).  At leading order and in the flat background, the change of $h_{ij}$ under diffeomorphisms takes the linear form  $\d_i \zeta_j + \d_j \zeta_i $ with $\zeta^i$ of order one (i.e., a function of the angles only).  Thus  we impose
\begin{eqnarray}
&& h_{ij} \equiv g_{ij} - \delta_{ij}  =   U_{ij} + j_{ij},  \label{eq:asymptgrav0a}\\
&& U_{ij} =
		\d_i \zeta_j + \d_j \zeta_i  = O(\frac{1}{r}),  \quad  \zeta^i = \zeta^i (\mathbf{n}^k) = O(1), \quad   \zeta^{i} (-\mathbf{n}^k) = \zeta^{i} (\mathbf{n}^k), \label{eq:asymptgrav0b}\\
&& j_{ij} = \frac{(\xbar h_{ij})^{even}(\mathbf{n}^k)}{r} + O(\frac{1}{r^2}), \quad (\xbar h_{ij})^{even} (-\mathbf{n}^k) = (\xbar h_{ij})^{even} (\mathbf{n}^k) \label{eq:asymptgrav0c}
\end{eqnarray}
for some $\zeta^i$ that may be assumed to be even since the odd parity component can be absorbed in a redefinition of $(\xbar h_{ij})^{even}$. 
Because $\zeta^i$ is of order one (the lower orders can be absorbed in a redefinition of the lower orders in $j_{ij}$, which are not restricted by parity conditions), one has
\be
U_{ij} = \frac{(\xbar h_{ij})^{odd}(\mathbf{n}^k)}{r}, \quad \Leftrightarrow \quad (\xbar h_{ij})^{odd}(\mathbf{n}^k) = r U_{ij}
\ee
with
\be
U_{ij}  = e^A_i \d_A \zeta_j + e^A_j \d_A \zeta_i
\ee

Similarly,   we allow both an odd component $(\xbar \pi^{ij})^{odd}(\mathbf{n}^k)$ and an even component $(\xbar \pi^{ij})^{even}(\mathbf{n}^k)$ in $\xbar \pi^{ij}$.  The odd component is unrestricted, but the even component must come from the transformation of $\pi^{ij}$ under diffeomorphisms (going to constants at infinity in order to preserve the asymptotic decay in $r^{-2}$). At leading order, $ \pi^{ij}$ sees only the normal diffeomorphisms.  Denoting the normal component of the diffeomorphism by $V$, with $V$ of order one, the transformation takes the form $\d^i \d^j V - \delta^{ij} \mathring \triangle V$ at leading order.   Here $\mathring \triangle \equiv \mathring \nabla ^i \mathring \nabla_i $ is the flat metric Laplacian.  Thus we take
\begin{eqnarray}
&& \pi^{ij} =V^{ij} + p^{ij}  \label{eq:asymptgrav0d}\\
&& V^{ij} = \d^i \d^j V - \delta^{ij} \mathring \triangle V, \quad V= V(\mathbf{n}^k) = O(1), \quad   V (-\mathbf{n}^k) = V (\mathbf{n}^k) \label{eq:asymptgrav0e} \\
&& p^{ij} = \frac{(\xbar \pi^{ij})^{odd}(\mathbf{n}^k)}{r^2} + O(\frac{1}{r^3}), \quad  (\xbar \pi^{ij})^{odd} (-\mathbf{n}^k) = - (\xbar \pi^{ij})^{odd} (\mathbf{n}^k) \label{eq:asymptgrav0f}
\end{eqnarray}
for some $V$ that may be assumed to be even since the odd parity components can be absorbed in a redefinition of $(\xbar \pi^{ij})^{odd}$.  
Because $V$ is of order one (the lower orders can be absorbed in a redefinition of the lower orders in $p^{ij}$, which are not restricted by parity conditions), one has
\be
V^{ij} = \frac{(\xbar \pi^{ij})^{even}(\mathbf{n}^k)}{r^{2}} \quad
\Leftrightarrow  \quad (\xbar \pi^{ij})^{even}(\mathbf{n}^k) = r^2
\delta^{im} \delta^{jn} V_{mn},
\ee
with
\begin{eqnarray}
 V_{mn} &=& \partial_m \partial_n V - \delta_{mn} \mathring \triangle V  \nonumber \\
 &=& - n_m n_n \frac{1}{r^2} \xbar \triangle V - \frac{1}{r} \Big(n_m {e^A}_n + n_n {e^A}_m \Big)\xbar D_A V  \nonumber \\
 &&  \qquad \qquad + {e^A}_m \, {e^B}_n \Big(\xbar D_A \xbar D_B V - \xbar \gamma_{AB} \xbar  \triangle V \Big)
\end{eqnarray}
where $\xbar \triangle \equiv \xbar \gamma^{AB} \xbar D_A \xbar D_B$ is the Laplacian on the sphere (with metric $\xbar \gamma_{AB}$).

Because the transformations linearize at infinity, the finite forms of the improper gauge transformations $U_{ij}$ and $V^{ij}$ coincide with their infinitesimal forms.  We can therefore assume that $\zeta^i$ and $V$ are finite, and not just infinitesimal, in the above formulas.

Adding these improper gauge transformation components to the dynamical
variables is not entirely trivial.  It complicates the form of the charges
and  undoes  the good property of finiteness of the symplectic term in the
action. To eliminate the divergences in the symplectic form while keeping
the freedom of performing improper gauge transformations,   we impose that
the constraints are fulfilled to leading orders.   For generic decays, the
constraints $\mathcal H^{grav} $ and $\mathcal H^{grav}_i $ typically
behave as $r^{-3}$ in cartesian coordinates.  We require them to hold two
orders lower, which yields
\be
\mathcal H^{grav}= O(r^{-5}), \quad \mathcal H^{grav}_i = O(r^{-5})  \quad \hbox{(in cartesian coordinates)}. \label{eq:asymptgrav1}
\ee
In fact, finiteness of the symplectic term and of the Hamiltonian requires only $\mathcal H^{grav}= O(r^{-4}) $ and $\mathcal H^{grav}_i = O(r^{-4})$.  We impose the stronger condition (\ref{eq:asymptgrav1}) in order to guarantee that the bulk pieces of the boost generators and of the rotation generators, which behave typically as $r \mathcal H^{grav}$ and $r \mathcal H^{grav}_i$, are also convergent integrals.

It is useful to write explicitly the condition that there is no $O(r^{-3})$-terms in the constraints. Denoting 
$J_{ij} = \frac{ (\xbar h_{ij})^{even}}{r}$ (first term in the expansion of $j_{ij}$) and $P^{ij} = \frac{(\xbar \pi^{ij})^{odd}}{r^2}$ (first term in the expansion of $p^{ij}$), one gets
\be
\d^i \d^j J_{ij} - \mathring \triangle J = 0, \quad \d_i P^{ij} = 0, \label{eq:asymptgrav100}
\ee
a condition that does not involve the improper gauge components since these automatically fulfill $\d^i \d^j U_{ij} - \mathring \triangle U = 0, \quad \d_i V^{ij} = 0$.  We shall verify below that this makes the symplectic form finite.  The next term in the expansion of the constraints can be found in \cite{Henneaux:2018cst}.

Imposing that the constraints hold at infinity two orders more strongly than what follows from the decay of the fields is a consistent request (the constraints transform into each other under Poincar\'e transformations) that does not eliminate any solution, for which the constraints hold throughout space, i.e., to all orders.

The second problem (absence of canonical generator for the boosts) is solved by demanding that the leading order of the mixed radial-angular components of the metric, which would be of order $O(1)$, be actually zero, so that
\be
g_{rA} \equiv h_{rA} = O(\frac{1}{r})
\ee
This implies\footnote{The condition $\mathring \nabla_r \zeta_A + \mathring \nabla_A \zeta_r = O(\frac{1}{r})$ is equivalent to the strict equation $\mathring \nabla_r \zeta_A + \mathring \nabla_A \zeta_r = 0$ because $\zeta_r$ depends only on the angles, $\zeta_r = U(x^A)$ and $\zeta_A$ is linear in $r$, $\zeta_A = r \xbar \zeta_A$ ($\zeta_i$ is assumed to be homogeneous since its lower order terms in $\frac{1}{r}$ are absorbed in $j_{ij}$.  Therefore, $\zeta_i$ does not depend on $r$ but only on the angles.)  The condition $\mathring \nabla_r \zeta_A + \mathring \nabla_A \zeta_r = 0$ yields by explicit computation $\xbar \zeta_A = \d_A U$.  Thus one has $\zeta_r = \d_r(r U)$ and $\zeta_A = \d_A(r U)$, i.e., $\zeta_i= \d_i (r U)$. }
\be 
\zeta_i = \d_i \tilde U, \quad \tilde U = r \xbar U, \quad \xbar U = \xbar U(\mathbf{n}^k) = O(1), \quad \xbar U (-\mathbf{n}^k) = -\xbar U (\mathbf{n}^k) \label{eq:asymptgrav2a}
\ee
for some function $\xbar U$ of the angles, and
\be
\xbar j_{rA} = 0. \label{eq:asymptgrav2b}
\ee
We shall also explicitly verify below that this solves the second difficulty.

The improper transformations fulfill interesting identities, which are
\be
 n^n U_{nm} = 0 \label{eq:0Identity1}
 \ee
and
\be 
r^2 n^m \, n^n V_{mn} = - \xbar \triangle V = r \xbar D^A (n^m e_A^n V_{mn}) = \xbar \gamma^{AB} e_A^m\, e_B^n V_{mn}. \label{eq:0Identity2}
\ee
To prove (\ref{eq:0Identity1}), we note that with $\zeta_i$ given by
$\partial_i(r\xbar U) = n_i \xbar U + r e^A_i \d_A \xbar U$, the quantity $U_{ij}$ becomes
\be
U_{ij} =    r e^A_i e^B_j \xbar \Upsilon_{AB}, \; \; \xbar \Upsilon_{AB} =
2(\xbar U \xbar \gamma_{AB}  + \xbar D_A \xbar D_B \xbar U).
\ee
One also sees from this expression that 
\be
\d_k U_{ij} =  (- n_k e^A_i e^B_j - n_i e^A_k e^B_j - n_j e^A_i e^B_ k) \xbar \Upsilon_{AB} + r e^A_i e^B_j e^C_k \xbar D_C \xbar \Upsilon_{AB} \label{eq:SecondIdenU}
\ee

The conditions (\ref{eq:asymptgrav0a})-(\ref{eq:asymptgrav0f}), (\ref{eq:asymptgrav1}) and (\ref{eq:asymptgrav2a})-(\ref{eq:asymptgrav2b}) are our set of boundary conditions.

\section{Finiteness of the symplectic structure}

We now verify the finiteness of the Hamiltonian kinetic term $\int d^3x \pi^{ij} \d_t g_{ij}$, i.e., of the symplectic structure.

There are four terms: $\int d^3x V^{ij} \d_t U_{ij}$, $\int d^3x p^{ij} \d_t U_{ij}$, $\int d^3x V^{ij} \d_t j_{ij}$ and $\int d^3x p^{ij} \d_t j_{ij}$.  The first term and the last terms are finite since the coefficient of the potential logarithmic divergence is in both case the integral over the sphere of an odd function and is therefore zero.  We thus need to study only the two terms $\int d^3x p^{ij} \d_t U_{ij}$ and $\int d^3x V^{ij} \d_t j_{ij}$.

The potential divergence in   $ \int d^3x p^{ij} \d_t U_{ij} $  is
$
\int d^3x P^{ij} \d_t U_{ij} 
$
with $P^{ij} = \frac{(\xbar \pi^{ij})^{odd}}{r^2}$.  Using the form of $U_{ij}$, and using the notation $\eta^i = \d_t \zeta^i$, this becomes 
\begin{eqnarray}
\int d^3x P^{ij} \d_t U_{ij} &=& 2 \int d^3 x P^{ij} \d_i \eta_j \nonumber \\
&=& 2 \int d^3x \d_i (P^{ij} \eta_j ) - 2 \int d^3 x \d_i (P^{ij}) \,  \eta_j \nonumber
\end{eqnarray}
The first term in the right hand side of the second line can be converted to a surface term, easily verified to be finite.  The second term is equal to zero thanks to (\ref{eq:asymptgrav100}).

Similarly, the potential divergence in $\int d^3x V^{ij} \d_t j_{ij}$ is
$
\int d^3x V^{ij} H_{ij} 
$
with $H_{ij} = \frac{\d_t (\xbar h_{ij})^{even}}{r}$.
Using the form of $V^{ij}$,  this becomes 
\begin{eqnarray}
\int d^3x V^{ij} H_{ij} &=& 2 \int d^3 x ( \d^i \d^j V - \delta^{ij} \mathring \triangle V) H_{ij} \nonumber \\
&=&\hbox{Finite surface terms} +   \int d^3x V (\d^i \d^j H_{ij} - \mathring \triangle H) \nonumber
\end{eqnarray}
The second term vanishes by (\ref{eq:asymptgrav100}) so that the potentially logarithmically divergent integral $\int d^3x V^{ij} \d_t j_{ij}$ is actually also finite.

We close this section by noting that the argument also shows that the improper gauge term $\int d^3x V^{ij} \d_t U_{ij}$ is in fact finite without having to assume definite parities for $U_{ij}$ or $V_{ij}$.  This can be for instance seen  by taking $j_{ij}$ of the same form as $U_{ij}$ (but with a $\zeta_i$ of opposite parity), and $p^{ij}$ of the same form as $V^{ij}$ (but with a $V$ of opposite parity).  Since these terms fulfill the constraints to leading order, the proof goes through.

\section{Invariance of the boundary conditions}

Under a deformation of the constant time hypersurface parametrized by $(\xi^\perp \equiv \xi, \xi^i)$,  the canonical variables transform as \cite{Dirac:1958sc,Arnowitt:1962hi}
\begin{eqnarray}
\delta g_{ij} &=& 2 \xi g^{-\frac12} \left(\pi_{ij} - \frac12 g_{ij} \pi \right) +\cL_{\xi} g_{ij} \label{eq:Transf0gij}\\
\delta \pi^{ij} &=& - \xi g^{\frac12}\left(R^{ij} - \frac12 g^{ij} R \right) + \frac12 \xi g^{-\frac12} \left(\pi_{mn} \pi^{mn} - \frac12 \pi^2 \right)  \nonumber \\
&& -2 \xi g^{-\frac12} \left(\pi^{im} {\pi_{m}}^j - \frac12 \pi^{ij} \pi \right) + g^{\frac12} \left(\xi^{\vert ij} - g^{ij} {\xi^{\vert m}}_{\vert m} \right) \nonumber \\
&& + \cL_{\xi} \pi^{ij} \label{eq:Transf0piij}
\end{eqnarray}
where $\cL_\xi g_{ij}$ and $\cL_\xi \pi^{ij}$ are respectively the Lie derivatives of $g_{ij}$ and $\pi^{ij}$ along the vector field $\xi^{i}$,
\begin{eqnarray}
\cL_\xi g_{ij} &=& \xi_{i \vert j} + \xi_{j \vert i} \\
\cL_\xi g_{ij} &=& \left(\pi^{ij} \xi^m \right)_{\vert m} - {\xi^i}_{\vert m} \pi^{mj} - {\xi^j}_{\vert m} \pi^{im}
\end{eqnarray}
and the vertical bar $\vert$ denotes the covariant derivative with respect to the full metric $g_{ij}$, i.e., $_{\vert i} \equiv \nabla_i$. 

The boundary conditions are invariant under hypersurface deformations  that behave asymptotically as 
\begin{eqnarray}
&& \xi = b_i x^i + T(\mathbf{n}) + C^{(b)}(\mathbf{n}) + O\left(r^{-1}\right) \label{eq:As1}\\
&&  \xi^i ={ b^i}_j x^j + I^i(\mathbf{n})+ O\left(r^{-1}\right) \label{eq:As2} \\
&& I_i(\mathbf{n}) = W_i(\mathbf{n}) + C^{(b)}_i (\mathbf{n}), \quad  W_i(\mathbf{n}) =  \d_i (r W(\mathbf{n})) \label{eq:As3}
\end{eqnarray}
where $b_i$ and $b_{ij} = -b_{ji}$ are arbitrary constants while
$T(\mathbf{n})$ and $W(\mathbf{n})$ are arbitrary functions on the unit
sphere ($\mathbf{n}^i = \frac{x^i}{r}$).   The constants $b_i$ parametrize the Lorentz
boosts (the corresponding term $- b^i x^0$ in $\xi^i$ can be absorbed in $a^i$
at any given time), whereas the antisymmetric constants $b_{ij} = -b_{ji}$
parametrize the spatial rotations.  The zero modes $a_0$ and $a_0^i$ of $T$
and $W^i$ are standard translations.  In a spherical harmonics expansion of $T(x^A)$ and $W(x^B)$, this corresponds to the choices $T(x^B) \sim a_0 Y^{0}_{0} $ and $W(x^B)  \sim a_0^m Y^1_m$ but higher spherical harmonics are allowed.  

The ``correcting terms'' $C^{(b)}(\mathbf{n})$ and $C^{(b)}_i (\mathbf{n})$ appear only when the transformation involves a boost ($b_i \not=0$). The first one is necessary for integrability of the boost charges.  The second  is necessary to maintain the condition $\xbar h_{rA} = 0$.  Their explicit form will be given in spherical coordinates in Subsection \ref{sec:Integrability}.  Both involve the fields and are of order one.   They are linear in the boost parameters so that $C^{(b)}(\mathbf{n}) = 0 = C^{(b)}_i (\mathbf{n})$ when $b_i= 0$.

The independent parameters of the symmetries are accordingly $b_i$, $b_{ij}$, $T(\mathbf{n})$ and $W(\mathbf{n})$.  The boundary conditions are therefore invariant under an asymptotic algebra that has the Poincar\'e algebra as a proper subalgebra.   General functions $T$ and $W$ describe ``angle-dependent'' translations, or ``supertranslations''. They can involve components of both parities.  The odd part of $W$ shifts $U$, while its even part modifies $h_{ij}$.  Similarly, the even part of $T$ shifts $V$ while its odd part modifies $p^{ij}$. The precise transformation rules of $U$ and $V$ are given in Appendix \ref{app:AsymTransfFields}.  It turns out that only the odd part of $W$ and the even part of $T$ define improper gauge transformations.  The other parts are pure gauge transformations (see next section).

\section{Surface terms and charge-generators}

The generator of the transformations (\ref{eq:As1})-(\ref{eq:As3}) is
\begin{equation}
	\label{eq:bms4gravgen}
	P^{grav}_{\xi}[g_{ij}, \pi^{ij}] = \int d^3x \, \left(\xi \mathcal H + \xi^i \mathcal H_i
	\right) + \mathcal B^{grav}_\xi[g_{ij}, \pi^{ij}],
\end{equation}
where the boundary term $\mathcal B_\xi$ is determined by the method of \cite{Regge:1974zd}, i.e., must be such that the exterior derivative $d_V P^{grav}_{\xi}$ of $P^{grav}_{\xi}$ in field space (with $(\xi, \xi^k)$ giben by (\ref{eq:As1})-(\ref{eq:As3})) reduces to a bulk integral involving only undifferentiated field variations $d_V g_{ij}$, $d_V \pi^{ij}$.  In other words, $d_V \mathcal B^{grav}_\xi$ must cancel the boundary terms generated from $d_V \int d^3x \, \left(\xi \mathcal H + \xi^i \mathcal H_i
	\right)$ through the integrations by parts necessary to bring $d_V P^{grav}_{\xi}$ to the appropriate bulk form. Given that the symplectic form $\Omega$ has the standard bulk expression $\int d^3x \, d_V \pi^{ij} \, d_V g_{ij}$ without surface term contributions, this is equivalent to requesting that the transformations (\ref{eq:As1})-(\ref{eq:As3}) be canonical transformations, i.e., leave the symplectic form invariant, $d_V i_\xi \Omega = 0$ ($\Leftrightarrow i_\xi \Omega = - d_V P^{grav}_{\xi}$, see \cite{Henneaux:2018gfi}).

Explicit computations lead to the identity \cite{Regge:1974zd}

\be
d_V \int d^3x \, \left(\xi \mathcal H + \xi^i \mathcal H_i
	\right) = \hbox{Requested bulk term} + \mathcal K_\xi[d_V
	g_{ij}, d_V \pi^{ij}]
\ee
where
\begin{eqnarray} \mathcal K_\xi[d_V
	g_{ij}, d_V \pi^{ij}] &=&  - \oint G^{ijkl} (\xi (d_V  g_{ij})_{\vert k} - \xi_{,k} d_V g_{ij}) d^2 S_l \nonumber \\
	&& \hspace{1cm} - \oint \left(2 \xi_k d_V \pi^{kl} + (2 \xi^k \pi^{jl} - \xi^l \pi^{jk}) d_V g_{jk} \right) d^2 S_l
	\label{eq:FormOfKappa}
	\end{eqnarray}
Here, $G^{ijkl} $ is De Witt supermetric,
\be
G^{ijkl}  = \sqrt{g} \Big( \frac12 (g^{ik} g^{jl} + g^{il} g^{jk}) - g^{ij} g^{kl} \Big).
\ee 
The surface integral potentially diverges for boosts and spatial rotations, which grow linearly with $r$.  

\subsection{Divergences are absent}

For the boosts, the potential divergence reads
$ - \oint \mathring{G}^{ijkl} (\xi  (\mathring \nabla_k( d_V U_{ij}) - \xi_{,k} d_V U_{ij}) d^2 S_l $
because: (i) to leading  order, $G^{ijkl}$ coincides with the De Witt supermetric $\mathring{G}^{ijkl}$ for the flat metric and the covariant derivative $(d_V g_{ij })_{\vert k}$ coincides with the covariant derivative $\mathring \nabla_k( d_V g_{ij})$ in the flat background; and (ii) due to the fact that $\xi$ is parity-odd for boosts, only the leading parity-odd component $U_{ij}$ contributes.  The parity-even component drops.  Since the constraints are fulfilled to leading order and $\xi$ is a Killing vector, the parity-odd component actually does not contribute either: $U_{ij}$ also drops and the divergence is absent.  This is because the integrand is an exact $2$-form, which integrates to zero.  This can be verified directly by expressing the leading component of the constraint and the integral over the $2$-sphere in polar coordinates as in Subsection \ref{sec:Integrability} below, or by observing that
$$ \oint \mathring{G}^{ijkl} (\xi  (\d_k( d_V U_{ij}) - \xi_{,k} d_V U_{ij}) d^2 S_l = \int d^3x \d_l \Big(\mathring{G}^{ijkl} ( \xi  \d_k( d_V U_{ij}) - \xi_{,k} d_V U_{ij}\Big)= 0$$
($\mathring \nabla_k = \d_k$ in cartesian coordinates) because the integrand $\d_l \Big(\mathring{G}^{ijkl} ( \xi  \d_k( d_V U_{ij}) - \xi_{,k} d_V U_{ij}\Big)$ vanishes on account of $\d_k \d_l \xi = 0$ and $\mathring{G}^{ijkl}  \d_k \d_l U_{ij} = \d^i \d^j U_{ij} - \mathring \triangle U = 0$.  So there is no divergence in the boost surface integral.

Similarly, the potential divergence in the angular momentum reads
$2 \oint  \xi^m \delta_{km} d_V V^{kl}   d^2 S_l $
where only the even-component of the leading part of $\pi^{ij}$ remains due to the fact that $\xi^k$ is odd for spatial rotations.  But again, because the constraints are satisfied to leading order, this is actually zero as integral over a closed $2$-surface of an exact $2$-form, or equivalently,
$$ 2 \oint  \xi_k  d_V V^{kl}   d^2 S_l = 2 \int d^3x \d_l \Big(  \xi_k  d_V V^{kl} \Big) = 0$$
as can be seen by using $\d_l \xi_k + \d_k \xi_l = 0$ and $\d_l V^{kl} = 0$. 

We stress that the argument needs only that the odd part of $\xbar h_{ij}$ and the even part of $\xbar \pi^{ij}$ fulfill the constraints to leading order.  In our case where these parts are given by improper gauge transformations, this is automatic and therefore, in fact, not an independent requirement.  For the alternative boundary conditions given in \cite{Henneaux:2018cst}, this is not automatic, however, and must be imposed.  [That the constraints hold asymptotically is in any case used in proving finiteness of the symplectic form with either sets of boundary conditions.] 

\subsection{Linear momentum and angular momentum}

Having established that the surface integral at infinity is finite, we must now verify that it is exact in field space (``integrable''), i.e., that it can be written as the exterior derivative $ - d_V \mathcal B^{grav}_\xi$ of a surface term $\mathcal B^{grav}_\xi$ that we would then add to the bulk term $ \int d^3x \, \left(\xi \mathcal H + \xi^i \mathcal H_i \right) $.

The analysis of the boosts is a bit cumbersome and most easily carried out in spherical coordinates.
By contrast, the surface terms for angular momentum and supertanslations are straightforwardly determined already in cartesian coordinates.  We start with them. Boosts are considered after the relevant formulas are rewritten in spherical coordinates.

\subsubsection*{Angular momentum}

For the angular momentum ($\xi = 0$, $\xi^m = {b^m}_k \, x^k = O(r)$), the $O(1)$-contribution to the above surface term is
\be
- 2 \oint  \xi^m \delta_{mk} d_V \Pi^{(3) kl} dS_l - 2 \oint \xi^m \left(h^{(1)}_{mk} d_V \Pi^{(2) kl} + \Pi^{(2) kl }d_V h^{(1)}_{mk} \right) d^2 S_l
\ee
where we have used the fact that $\xi^l dS_l = 0$, and where $h^{(1)}_{mk}$ is the $O(r^{-1})$ component of $h_{ij}$,
$$h_{ij} = h^{(1)}_{mk} + O(r^{-2}), \quad  h^{(1)}_{mk} = \frac{\xbar h_{mk} (\mathbf{n})}{r}$$  
while $\Pi^{(2) kl}$ and  $\Pi^{(3) kl}$ are respectively the $O(r^{-2})$ and $O(r^{-3})$ components of $\pi^{kl}$ (in cartesian coordinates),
$$\pi^{kl} = \Pi^{(2) kl} + \Pi^{(3) kl} + O(r^{-4}), \quad \Pi^{(2) kl} =  \frac{\xbar \pi^{ kl} (\mathbf{n})}{r^2}, \quad \quad \Pi^{(3) kl} =  \frac{\xbar \Pi^{(3)kl} (\mathbf{n})}{r^3}.$$
The surface term is clearly an exact form in field space,
\be
d_V \Big[ - 2 \oint \xi^m \Big(\delta_{mk}  \Pi^{(3) kl} + h^{(1)}_{mk}  \Pi^{(2) kl} \Big) dS_l \Big]
\ee
Therefore, adjusting the integration constant so that Minkowski space has zero angular momentum, one finds that the surface term to be added to$\int d^3x \,  \xi^i \mathcal H_i$  for rotations is 
\be
\frac 12 b_{mn} M^{mn}, \quad M^{mn} = 4\oint x^{[n}\Big(\ \Pi^{(3) m]l} + \delta^{m]p}h^{(1)}_{pk}  \Pi^{(2) kl} \Big) dS_l
\ee
(with antisymmetrization of weight one).

\subsubsection*{Supertranslations}

We now consider supertranslations, generated by $\xi = T$ and $\xi_i = \partial_i (\tilde W)$, $\tilde W = r W$.  The $O(1)$ piece of the surface integral at infinity reads then
$$ -\oint \mathring G^{ijkl} T (d_V  h^{(1)}_{ij})_{, k} d^2 S_l - \oint \left(2 \xi_k d_V \Pi^{(2) kl}  \right) d^2 S_l $$
and is clearly integrable since it is linear in the $d_V$ of the fields.  We have dropped the term $\oint \mathring G^{ijkl} (T_{,k} d_V h^{(1)}_{ij}) d^2 S_l$  which is zero since $T$ depends only on the angles and hence $T_{,k} \sim e^A_k$, while $d^2 S_l \sim n_l$ so that only $d_V h^{(1)}_{rA}$ appears in that expression.  But $h^{(1)}_{rA} = 0$.

It turns out that the improper gauge parts of  $  h^{(1)}_{ij}$ and $ \Pi^{(2) kl} $ do not contribute to the integral, so that the surface term to be added to $\int d^3x \, \left(T \mathcal H + \partial^i (\tilde W) \mathcal H_i \right) $ reduces to
$$  \mathcal B^{grav}_{\{T, W\}} = \oint \mathring G^{ijkl} \, T  \,  \Big((h^{ (1)}_{ij})^{even}\Big)_{, k}  d^2 S_l + \oint \left(2 \d_k(\tilde W)   (\Pi^{ (2) kl} )^{odd} \right) d^2 S_l $$
(with $\mathcal B^{grav}_{\{T, W\}} =0$ for $h_{ij} = \pi^{ij} = 0$).  It follows that only $T^{even}$ and $W^{odd}$ appear in the surface integral.

That the improper gauge part of  $  h^{(1)}_{ij}$ indeed drops from the surface integral follows from the fact that $d^2 S_l \sim n_l$ and $\mathring G^{ijkl} \, n_l \, U_{ij, k}= 0$ due to the form (\ref{eq:SecondIdenU}) of $U_{ij}$ and the identity
$ \mathring G^{ijkl} \, n_l (n_k e^A_i e^B_j + n_i e^A_k e^B_j + n_j e^A_i e^B_k ) = 0. $
That the improper gauge part of $ \Pi^{(2) kl} $ also drops is a consequence of (\ref{eq:0Identity2}), of the relation   $\partial_k (r W) = n_k W + e^A_k \d_A W$ and of the fact that the integral $\oint \partial_A V^A d^2 x$ of a total derivative on the sphere vanishes.

The generator of supertranslations is thus given by
\be 
Q_{ \{T,W\}} = \int d^3x \, \left(T \mathcal H + \partial^i (\tilde W) \mathcal H_i \right) + \mathcal B^{grav}_{\{T, W\}}.
\ee
When the constraints hold, $Q_{ \{T,W\}} \approx \mathcal B^{grav}_{\{T, W\}}$, an expression that vanishes when $T$ is odd and $W$ is even.   This means that the transformations generated by $T^{odd}$ and $W^{even}$ are proper gauge transformations that do not change the physical state of the system.  By contrast, the transformations generated by $T^{even}$ and $W^{odd}$ have generically non-vanishing charges and are truly improper gauge transformations.  For $T=1$ and $W = 0$, we recover the ADM energy \cite{Arnowitt:1962hi}, \cite{Regge:1974zd}. 

The expression of the charge associated with supertranslations $T$ in the normal direction to the constant time hypersurfaces can be simplified by using the property
$(h^{ (1)}_{ij})^{even} = \frac{1}{r} n_i n_j \xbar h_{rr} + r e^A_i e^B_j \xbar h_{AB}$
where $\xbar h_{rr}$ and $\xbar h_{AB}$ depend only on the angles. The $\xbar h_{AB}$ term drops exactly as the $U_{ij}$ term, and one is left with $\xbar h_{rr}$ only.  This yields
\be
\mathcal B^{grav}_{\{T, 0\}}= 2 \oint d^2 x \, \sqrt{\xbar \gamma} \,  T \, \xbar h_{rr}
\ee

\subsubsection*{Lapse and shift}
The lapse $N$ and the shift $N^k$, which are Lagrange multipliers for the (first-class) constraints,  must be chosen so that the dynamical evolution preserves the boundary conditions.  This means that they can be taken to parametrize a generic asymptotic symmetry.  It is customary to take:
\begin{equation}
	\label{eq:asymptgravII}
	N = 1 + O(r^{-1}), \quad N^r = O(r^{-1}), \quad N^A = O(r^{-2}).
\end{equation}
This corresponds to slicings by hypersurfaces that become asymptotically parallel hyperplanes. Imposing these boundary conditions on the lapse and the shift implies that we have to add to the action  the ADM energy, i.e, 
\begin{equation}
B_{\infty} = \mathcal B^{grav}_{\{1, 0\}} = \oint d^2x
	\sqrt{\xbar \gamma} \,2 \xbar h_{rr}.
	\label{eq:BounTermAction0}
\end{equation}

\subsection{Spherical coordinates}
\label{sec:sphericalcoords}

In order to analyze integrability of the boost charges, we go to spherical coordinates. 

Recalling that the conjugate momenta carry a density weight, one finds that the above boundary conditions on the dynamical variables read in spherical coordinates:
\begin{equation}
	\label{eq:asymptgravI}
	\begin{array}{rclrcl}
		g_{rr} &=& 1 + \frac{1}{r} \xbar h_{rr} + \frac{1}{r^2} h^{(2)}_{rr} +
		o(r^{-2}),&
		\pi^{rr} &=& \xbar \pi^{rr} + \frac{1}{r} \pi^{(2)rr} +
		o(r^{-1}),\\
		g_{rA} &= & \xbar \lambda_A + \frac{1}{r} h^{(2)}_{rA} +
		o(r^{-1}), (\xbar \lambda_A = 0), &
		\pi^{rA} & = & \frac{1}{r} \xbar \pi^{rA} + \frac{1}{r^2} \pi^{(2)rA} +
	o(r^{-2}),\\
	g_{AB} & = & r^2 \xbar\gamma_{AB} + r \xbar h_{AB} +  h^{(2)}_{AB} +
	o(1),& 
	\pi^{AB} & = & \frac{1}{r^2} \xbar \pi^{AB} + \frac{1}{r^3} \pi^{(2)AB}+
	o(r^{-3}).
\end{array}
\end{equation}
Although it is absent, we have allowed  here a term of order $1$ in $g_{rA}$ parametrized by $\xbar h_{rA} \equiv \xbar \lambda_A$ in order to show that  integrability of the boost charges is indeed guaranteed if one imposes  $\xbar \lambda_A = 0$. 
Note that what we call here $\pi^{(2)rr}$, $\pi^{(2)rA}$, $\pi^{(2)AB}$ (second term in the expansion of the momenta) was denoted $\Pi^{(3)ij}$ before (terms of order $r^{-3}$ in cartesian coordinates). 

As we have seen, the leading orders of the dynamical variables are further subject to two types of extra conditions:
(i) Parity conditions; (ii)  Constraint conditions.  

\subsubsection*{Generalized Parity conditions}

Under the antipodal map $x^A \rightarrow -x^A$, the above parity conditions involving an improper gauge twist are
\begin{gather}
	\label{eq:ParitySpher001}
	\xbar h_{rr} = \text{even}, \quad \xbar \lambda_A = \text{odd}, \\
	 \xbar \pi^{rr}  = (\xbar \pi^{rr})^{odd} - \sqrt{\xbar\gamma} \, \xbar \triangle V
	, \quad \xbar \pi^{rA} =
	(\pi^{rA})^{even} - \sqrt{\xbar\gamma}\, \xbar D^A V, \\
	\xbar \pi^{AB} = (\xbar \pi^{AB})^{odd} + \sqrt{\xbar\gamma}(\xbar D^A \xbar D^B V- \xbar
	\gamma^{AB} \xbar \triangle V), \\ \xbar h_{AB} =
	(\xbar h_{AB})^{even} + 2(\xbar D_A \xbar D_B U + U \xbar \gamma_{AB})
	\label{eq:pParitySpher002}
\end{gather}
with $V$ even and $U$ odd ($\xbar \triangle \equiv \xbar D_C \xbar D^C$). It follows from these conditions that 
\be \xbar \pi^{rr} - \xbar \pi^A_A = \text{odd} \; \; \text{ (strictly). }
\ee 

The variables $\xbar h_{rr}$ and $\xbar \pi^{rr} - \xbar \pi^A_A$ play thus a special role: they are invariant under improper gauge transformations, that is, they are invariant under supertranslations.

\subsubsection*{Constraint conditions}

In order to write down the constraint conditions in polar coordinates,  it is useful to introduce a 1+2 radial split of the 3 dimensional metric $g_{ij}$:
\begin{gather}
	g_{rr} = \lambda^2 +  \gamma_{AB}\lambda^A \lambda^B, \quad g_{rA} =
	\gamma_{AB}\lambda^B, \quad g_{AB} = \gamma_{AB}, \\
	\lambda = 1 + r^{-1} \xbar \lambda +  r^{-2} \lambda^{(2)} +
	o(r^{-2}), \quad \lambda^A = r^{-2} \xbar \lambda^A + r^{-3} \lambda^{(2)A} + o(r^{-3}), \\
	\gamma_{AB} = r^2 \xbar \gamma_{AB} + r \xbar h_{AB} + h^{(2)}_{AB} +
	o(1).
\end{gather}
In a manner consistent with our previous conventions for barred quantities, we will use $D_A$ and $\xbar D_A$ to respectively denote the covariant derivatives associated with $\gamma_{AB}$ and $\xbar \gamma_{AB}$. The indices $A, B, ...$ on bulk fields will be lowered and raised with $\gamma_{AB}$ and its inverse $\gamma^{AB}$ while the same indices on asymptotic fields will be lowered and raised with $\xbar\gamma_{AB}$ and its inverse $\xbar \gamma^{AB}$.  
One has the dictionary
\be
\xbar \lambda = \frac12 \xbar h_{rr}, \quad \xbar \lambda^A = \xbar \gamma^{AB} \xbar h_{rB} \equiv \xbar \gamma^{AB} \xbar \lambda_B
\ee

The extrinsic curvature of the constant $r$ surfaces is given by
\begin{equation}
	K_{AB} = \frac{1}{2\lambda} (-\d_r \gamma_{AB} + D_A \lambda_B + D_B
	\lambda_A)
\end{equation}
and differs from the flat space $\mathring K_{AB}$ (which blows up like $r$) by correction terms of order $1$.    We will expand $K^A_B$, which behaves like $r^{-1}$ in flat space, as follows
\begin{equation}
	 K^A_B = -r^{-1} \delta^A_B + r^{-2} \xbar k^A_B +
	r^{-3} {k^{(2)}}^A_B + o(r^{-3})
\end{equation}
where the first term is the flat space value.  Note that this yields
\be
K_{AB} = - r \xbar \gamma_{AB} + (\xbar k_{AB} - \xbar h_{AB}) + o(1)
\ee
so that $\xbar k_{AB}$ is not the coefficient of the first subleading term in the expansion of $K_{AB}$ (it is defined through the expansion of $K^A_B$). A direct computation gives
\be
\xbar k_{AB} = \frac12 \xbar h_{AB} + \xbar \lambda \xbar \gamma_{AB} + \frac12 (\xbar D_A \xbar \lambda_B + \xbar D_B \xbar \lambda_A)
\ee
from which
the relationship
$\xbar k = \frac12 \, \xbar h + 2 \xbar \lambda + \xbar D_A \xbar \lambda^A
$
follows.
The (generalized) parity conditions imply
\begin{gather}
	\label{eq:ParitySpher1}
	\xbar \lambda = \text{even}, 
	 \\ \xbar k_{AB} =
	(\xbar k_{AB})^{even} + \xbar D_A \xbar D_B U + U \xbar \gamma_{AB}
	\label{eq:pParitySpher2}
\end{gather}

Various useful formulas on the $2+1$ decomposition of the curvature are given in Appendix \ref{app:2+1}.  A straightfoward derivation using these formulas shows then that
 the conditions  that the constraints should decrease faster than the anticipated rate are linear conditions on the asymptotic fields that read
 \begin{gather}
	\label{eq:asymptconst}
	\xbar D_A \xbar D_B \xbar k^{AB} -
	\xbar D_A \xbar D^A \xbar k = 0, \quad \xbar \pi^{rA} +\xbar D_B \xbar
	\pi^{AB} = 0,\quad \xbar D_A \xbar D_B \xbar
	\pi^{AB} + \xbar \pi^A_A = 0.
\end{gather}

\subsection{Integrability analyzed in spherical coordinates}
\label{sec:Integrability}

\subsubsection*{Asymptotic symmetry transformations}

The asymptotic symmetries preserving the boundary conditions are generated by the vector fields (\ref{eq:As1})-(\ref{eq:As3}), which read in spherical coordinates
\begin{gather}
	\label{eq:asymptsymI}
	\xi =  br  + T + C_{(b)} + O(r^{-1}),\quad \xi^A =
	Y^A + \frac{1}{r} \Big(\xbar D^A W + C^A_{(b)}
	\Big) + O(r^{-2}),\\
	\label{eq:asymptsymII}
	\xi^r = W + O(r^{-1}), \quad \xbar D_A \xbar D_B b + \xbar \gamma_{AB} b = 0, \quad \cL_Y \xbar
	\gamma_{AB} = 0, 
\end{gather}
where $b(x^B), Y^A(x^B)$ describe boosts and spatial rotations, while $T(x^B)$ and $W(x^B)$ are field-independent functions on the sphere.  One has $b_i x^i = b r$ while   $Y^A = \frac12 b^{mn} Y^A_{mn}$ in terms of the basis $x_m \d_{n} - x_n \d_m \equiv Y_{mn} $.

The ``correcting'' terms  $C_{(b)}$ and $C^A_{(b)}$ are linear in the boosts, depend on the fields and are of order one.  They are thus  subleading with respect to $r b$.   The term $C_{(b)}$ must be added to the boosts in order to make their generator well defined.  We write it as $C_{(b)} = bF$ with a function $F$ that will be determined through the integrability requirement to be equal to $- \xbar k$ (up to the irrelevant addition of a multiple of $\xbar \lambda$, see below), 
\be
C_{(b)} = bF, \qquad F =  - \xbar k \label{eq:ExprForF}
\ee

The term $C^A_{(b)}$ is also a correction term that accompanies the boosts and which is necessary to maintain the condition $\xbar h_{rA}= 0$.  A straightforward computation using the transformation of $g_{ij}$ under boosts shows that it is given by
\be
C^A_{(b)} = \frac{2b}{\sqrt{\xbar\gamma}}
	\xbar \pi^{rA}
\ee

Using the $2+1$ decomposition, the surface term (\ref{eq:FormOfKappa}) reads in polar coordinates
\begin{multline}
\mathcal K_\xi[d_V
	g_{ij}, d_V \pi^{ij}] = 
	\oint d^2x \, \Big \{ -2 \xi^i d_V \pi^r_i + \xi^r
		\pi^{ij} d_V g_{ij} - 2\sqrt \gamma \xi  d_V K\\
	- \sqrt \gamma \gamma^{BC} d_V \gamma_{CA}\left(\xi K^A_B + \frac{1}{\lambda} (\d_r \xi -
	\lambda^D \d_D \xi)\delta^A_B\right) \Big\},
\end{multline}
an expression that can be further transformed to
\begin{multline}
\mathcal K_\xi[d_V
	g_{ij}, d_V \pi^{ij}] = 
	r \oint d^2x \, \Big \{ -2 \, Y^A \xbar\gamma_{AB} d_V\xbar\pi^{rB}  -2\sqrt{\xbar \gamma} b d_V
		\xbar k\Big\} \\
			+\oint d^2x \Big\{-2 Y^A d_V (\xbar h_{AB} \xbar \pi^{rB} + \xbar \gamma_{AB}
	\pi^{(2)rB} + \xbar \lambda_A \xbar \pi^{rr}) -2 \Big(\xbar D^A W + C^A_{(b)}\Big)\xbar\gamma_{AB} d_V \xbar
	\pi^{rB} \\- 2 W d_V \xbar \pi^{rr}- \sqrt{\xbar\gamma} (b \xbar h d_V \xbar k + 2 (bF + T)  d_V
	\xbar k + 2b d_V k^{(2)})\\
	+\sqrt{\xbar \gamma} ( bF + T + \xbar \lambda b + \xbar
	\lambda^D\d_D b )d_V \xbar h-\sqrt{\xbar \gamma} 
	b\, \xbar k^{AB}d_V \xbar h_{AB}  \Big\}+ o(r^0).
\end{multline}

\subsubsection*{Divergences are absent - Another check}

The first line, proportional to $r$, is the potentially linearly divergent term identified above, the  coefficient of  which can be again checked from (\ref{eq:asymptconst}) to be actually zero.  Indeed one has
$$ \oint d^2x  \, Y^A \xbar\gamma_{AB} d_V\xbar\pi^{rB} =- \oint d^2x  \, Y^A \xbar\gamma_{AB} \xbar D_C d_V\xbar\pi^{BC} = \oint d^2x \xbar D_{(C} Y_{B)} d_V\xbar\pi^{BC}  = 0
$$
since $Y^A$ is a Killing vector on the sphere.  Similarly,
\begin{eqnarray}
 \oint \sqrt{\xbar \gamma} b d_V \xbar k &=& \oint \sqrt{\xbar \gamma} b d_V \Big( \xbar k - \xbar D_A \xbar D_B \xbar k^{AB} + \xbar D_A \xbar D^A \xbar k \Big) \nonumber \\
 &=& \oint \sqrt{\xbar \gamma} b d_V \Big( \xbar k + \xbar k  - 2 \xbar k \Big) = 0
\end{eqnarray}
upon integration by parts and use of the fact that $b$ fulfills $\xbar D_A \xbar D_B b + \xbar \gamma_{AB} b = 0$.

Strict parity conditions are therefore not necessary to eliminate the linear divergence in $\mathcal K_\xi[d_V
	g_{ij}, d_V \pi^{ij}]$.  One may allow a twist in the parity conditions given by an improper gauge transformation, which automatically solves the constraints to leading order.
	
	\subsubsection*{Angular momentum and supertranlation charges in polar coordinates}

The terms proportional to $Y^A$ in the finite piece of $\mathcal K_\xi[d_V g_{ij}, d_V \pi^{ij}]$ leads to the angular momentum constructed previously,
\be
	\frac 12 b_{mn} M^{mn} = \oint d^2x \Big
	\{  
	Y^A \Big(4\xbar k_{AB} \xbar \pi^{rB} - 4 \xbar \lambda\xbar \gamma_{AB}
	\xbar\pi^{rB}+ 2 \xbar \gamma_{AB}
	\pi^{(2)rB}\Big)
\ee
where we have set $\xbar \lambda_A = 0$ to anticipate the result of the analysis of the boosts.  This term could be kept here since it does not spoil integrability of the angular momentum.  The terms proportional to $T$ and $W$ yields the supertanslation charges,
$
\mathcal B^{grav}_{\{T, W\}}=  \oint d^2x \Big
	\{   2 W \Big( \xbar \pi^{rr} -  \xbar \pi^A_A\Big)  + 4 \, T \sqrt{\xbar\gamma}\,  \xbar
	\lambda \Big\}
$
(where we used the asymptotic form of the momentum constraint) and can be simplified to
\be
\mathcal B^{grav}_{\{T, W\}}=  \oint d^2x \Big
	\{   2 W^{odd} \Big( \xbar \pi^{rr} -  \xbar \pi^A_A\Big)  + 4 \, T^{even} \sqrt{\xbar\gamma}\,  \xbar
	\lambda \Big\}
\ee
since $\xbar \pi^{rr} -  \xbar \pi^A_A$ is odd while $\xbar \lambda$ is even. These expressions are  in complete agreement with what we found before.

\subsubsection*{Boosts}
It remains to analyse the boosts, for which $\mathcal K_\xi[d_V
	g_{ij}, d_V \pi^{ij}] $ reduces to
\begin{multline}
\mathcal K_\xi[d_V
	g_{ij}, d_V \pi^{ij}] = 
			\oint d^2x \Big\{ -2  C^A_{(b)}\xbar\gamma_{AB} d_V \xbar
	\pi^{rB} - \sqrt{\xbar\gamma} (b \xbar h d_V \xbar k + 2 bF   d_V
	\xbar k + 2b d_V k^{(2)})\\
	+\sqrt{\xbar \gamma} ( bF  + \xbar \lambda b + \xbar
	\lambda^D\d_D b )d_V \xbar h-\sqrt{\xbar \gamma} 
	b\, \xbar k^{AB}d_V \xbar h_{AB}  \Big\}+ o(r^0).
\end{multline}
One can rewrite this term as
\begin{multline}
\mathcal K_\xi[d_V
	g_{ij}, d_V \pi^{ij}] =
	d_V\oint d^2x \, \Big \{	
	- 2\sqrt{\xbar\gamma} b k^{(2)}  - \sqrt{\xbar\gamma} \frac{1}{4} b
	( \xbar h^2 + \xbar h^{AB} \xbar h_{AB}) - b\frac{2}{\sqrt{\xbar\gamma}} \xbar \gamma_{AB}
	\xbar\pi^{rA}\xbar\pi^{rB}\Big\}\\
	+\oint d^2x \, \Big \{	
	- \sqrt{\xbar\gamma}(2bF +  \xbar h b) \, d_V   (2\xbar \lambda
	+ \xbar D_A \xbar \lambda^A)\\ + \sqrt{\xbar\gamma} (\xbar \lambda^C\d_C
	b \,\xbar\gamma^{AB} - b\xbar D^A \xbar\lambda^B) \, d_V   \xbar h_{AB}\Big\}+ o(r^0). \label{eq:KFull}
\end{multline}
The terms in the second and third lines of (\ref{eq:KFull}) are not integrable without restriction.  This can be seen by computing $d_V \mathcal K_\xi$.   One then first note that the term in  $ d_V \xbar h d_V \xbar \lambda$ in $d_V \mathcal K_\xi$  reads $b(2d_V F + d_V \xbar h) d_V \xbar \lambda$, and this is zero only if $F $ is taken to be equal to $- \frac 12 \xbar h + a \xbar \lambda$, where the constant $a$ is arbitrary.  This is the same as the expression (\ref{eq:ExprForF}) up to a term $a' \xbar \lambda$, but this integrable term plays no role because it yields  a vanishing integral $\oint d^2 x b (\xbar \lambda)^2 = 0$ since $b$ is odd while $ (\xbar \lambda)^2$ is even. We can thus take it as we please.  We have set $a' = 0$ for convenience. The condition $\xbar \lambda_A = 0$ makes the last line also integrable.   We have not studied in depth the general solution of the demand that the one-form in field space  $\mathcal K_\xi$ should be exact but as we shall discuss in the conclusions, the condition $\xbar \lambda_A = 0$ makes a lot of sense.  Further support is given in Appendix \ref{sec:IntegrabilityLambda}.

Taking into account these conditions, one finds explicitly for the surface term $\mathcal B_\xi$ associated with boosts (such that $d_V \mathcal B_\xi + \mathcal K_\xi = 0$)
\be
\label{eq:boundtermBoost}
    \mathcal B_\xi[g_{ij}, \pi^{ij}] = \oint d^2x \, \Big \{ b\,
	\sqrt{\xbar \gamma} \Big( 2  k^{(2)} + \xbar k^2 +
	\xbar k^A_B \xbar k^B_A - 6 \xbar\lambda\xbar k\Big) +b\frac{2}{\sqrt{\xbar\gamma}} \xbar \gamma_{AB}
    \xbar\pi^{rA}\xbar\pi^{rB})\Big\}.
\ee
We have again adjusted the integration constant in $\mathcal B_\xi$ so that the background Minkowski solution has zero charge.

\section{$BMS_4$ Algebra - Null infinity}

Putting all pieces together, one finds that the asymptotic symmetries are canonical transformations generated by
\begin{equation}
	\label{eq:bms4gravgen2}
	P^{grav}_{\xi}[g_{ij}, \pi^{ij}] = \int d^3x \, \left(\xi \mathcal H + \xi^i \mathcal H_i
	\right) + \mathcal B^{grav}_\xi[g_{ij}, \pi^{ij}].
\end{equation}
Here, the boundary term is given by the sum of the contributions found before
\begin{multline}
	\label{eq:boundtermBgen}
	\mathcal B_\xi[g_{ij}, \pi^{ij}] = \oint d^2x \Big
	\{  
	Y^A \Big(4\xbar k_{AB} \xbar \pi^{rB} - 4 \xbar \lambda\xbar \gamma_{AB}
	\xbar\pi^{rB}+ 2 \xbar \gamma_{AB}
	\pi^{(2)rB}\Big) +2 W \Big( \xbar \pi^{rr} -  \xbar \pi^A_A\Big) \\ + T\, 4\sqrt{\xbar\gamma}\,  \xbar
	\lambda +b\,
	\sqrt{\xbar \gamma} \Big( 2  k^{(2)} + \xbar k^2 +
	\xbar k^A_B \xbar k^B_A - 6 \xbar\lambda\xbar k\Big) +b\frac{2}{\sqrt{\xbar\gamma}} \xbar \gamma_{AB}
	\xbar\pi^{rA}\xbar\pi^{rB}\Big\}.
\end{multline}
Note that the boost charge and the angular momentum involve contributions that are quadratic in the asymptotic fields.  These are absent for the strict parity conditions  \cite{Regge:1974zd}.  By making a BMS transformation away from an ``untwisted frame'', one therefore generates quadratic contributions.

The algebra is easily evaluated to be:
\begin{equation}
	\Big\{P^{grav}_{\xi_1}[g_{ij}, \pi^{ij}], P^{grav}_{\xi_2}[g_{ij}, \pi^{ij}]\Big\} =
	P^{grav}_{\hat\xi}[g_{ij}, \pi^{ij}],
\end{equation}
where $\hat \xi$ generates an asymptotic symmetry with the following parameters
\begin{flalign}
	\label{eq:hamilbmsI}
	\hat Y^A & = Y^B_1\d_B Y_2^A + \xbar \gamma^{AB} b_1\d_B b_2 - (1
	\leftrightarrow 2),\\
	\label{eq:hamilbmsII}
	\hat b & = Y^B_1\d_B b_2 - (1 \leftrightarrow 2),\\
	\label{eq:hamilbmsIII}
	\hat T & = Y_1^A\d_A T_2 - 3 b_1 W_2 - \d_A b_1 \xbar D^A W_2 - b_1
	\xbar D_A\xbar D^A W_2 - (1 \leftrightarrow 2),\\
	\label{eq:hamilbmsIV}
	\hat W & = Y_1^A \d_A W_2 - b_1T_2 - (1 \leftrightarrow 2).
\end{flalign}

Since the boosts are characterized by odd functions $b$, we see that the
Lorentz transformations mix $T^{even}$ with $W^{odd}$, and $T^{odd}$ with
$W^{even}$.  That is, the pair $(T, W)$ does not form an irreducible
representation of the Lorentz group but breaks into separate Lorentz
multiplets $(T^{even}, W^{odd})$ and $(T^{odd}, W^{even})$. The multiplet
$(T^{even}, W^{odd})$ generates  physical (``improper'') gauge transformations and contains the translations.  The multiplet $(T^{odd}, W^{even})$ generates pure gauge transformations that do not change the physical state of the system.  The strict parity conditions eliminate the physical transformations $(T^{even}, W^{odd})$, which is not an acceptable gauge fixing. 

Modding out the trivial transformations generated by even $W$'s  and odd $T$'s, the resulting algebra is the algebra found in \cite{Henneaux:2018cst}. Using the results of \cite{Troessaert:2017jcm}, this algebra was shown there to be the $BMS_4$ algebra expressed in an unfamiliar parametrization. This was done by integrating the equations of motion for the symmetry parameters all the way to null infinity, along the lines of  \cite{Ashtekar:1978zz,BeigSchmidt,Beig:1983sw,Fried1,Friedrich:1999wk,Friedrich:1999ax}.  One finds that the odd $W$'s  and even $T$'s combine to yield the arbitrary function of the angles parametrizing supertranslations in the null infinity parametrization.  This enables one to conclude that the symmetry at spatial infinity is the same $BMS_4$ as the $BMS_4$ uncovered at null infinity.
We also note that our analysis implies the matching conditions  between the values of the fields on the future and past critical spheres (i.e., the past boundary of future null infinity and the future boundary of past null infinity) imposed in \cite{Strominger:2017zoo}.  We refer to \cite{Henneaux:2018cst,Henneaux:2018hdj,Troessaert:2017jcm} for the details.

\section{Conclusions}

We have reviewed here our work on the Hamiltonian formulation of Einstein gravity in the context of asymptotically flat spacetimes \cite{Henneaux:2018cst,Henneaux:2018gfi,Henneaux:2018hdj}.  The  boundary conditions that we have proposed at spatial infinity lead to a consistent dynamics invariant under the BMS group, which has a non trivial action.

Our boundary conditions are characterized by two key features.

\begin{itemize}
\item First, they involve parity conditions on the leading order of the fields, which differ from strict parity conditions by a twist equal to an improper gauge transformation.  These generalized parity conditions make the symplectic form and the Lorentz charges finite, which was the motivation for introducing them. 

There is another way to motivate the generalized parity conditions. The Weyl tensor of generic initial data not fulfilling any parity condition develop logarithmic singularities as one goes to null infinity, unless one imposes strict parity conditions on the Weyl tensor components \cite{Ashtekar:1990gc,Herberthson:1992gcz,Troessaert:2017jcm}.  These strict parity conditions can be formulated, at spatial infinity, as the conditions that the electric and magnetic components of the Weyl tensor, which are strictly invariant under proper and improper gauge transformations, be strictly even in cartesian coordinates.  Together with the parity conditions that the BMS invariant $\xbar \lambda$ (related to the mass) should be even and its BMS invariant conjugate $\xbar \pi^{rr}- \xbar \pi^{A}_{A}$ (related to the linear momentum) should be odd, these strict parity conditions on the Weyl tensor imply our generalized parity conditions \cite{Henneaux:2018hdj}\footnote{The fact that the matching conditions of \cite{Strominger:2017zoo} follow from the generalized parity conditions is yet another motivation.}.  

Similar features are present in electromagnetism \cite{Henneaux:2018gfi}, as well as for $2$-form gauge fields \cite{Henneaux:2018mgn}.  We note, however, that convergence of the symplectic form and absence of singularities in the physical fields do not impose parity restrictions in higher dimensions, at least for electromagnetism \cite{Henneaux:2019yqq}.  Preliminary investigations seem to indicate that this is also the case for gravity and it remains to be seen whether a consistent formulation free from parity conditions can  be given for higher dimensional Einstein theory. 

\item A second key ingredient is $\xbar h_{rA} = 0$, which expresses that  the expected leading order of the mixed radial-angular components of the metric should actually be zero. This condition is sufficient to make the boost charges integrable.  One way to motivate it is given by another important implication of this condition.  As we have seen, it relates the angular part $\xi^A$ of the improper diffeomorphisms to their radial part $\xi^r$. This is what makes the BMS group emerges at spatial infinity.  A similar relationship between $\xi^r$ and $\xi^A$ plays also a central role in the AdS context, where the analog of $\xbar h_{rA} = 0$ is also naturally imposed \cite{Henneaux:1985tv,Brown:1986nw}.  In appendix \ref{sec:IntegrabilityLambda}, we  further study  the necessity of the condition $\xbar h_{rA} = 0$.  First, we investigate in greater detail the integrability  of  the expression (\ref{eq:KFull}) contributing to the  boost charges with the canonical symplectic structure adopted throughout the paper, and show that while more general $\xbar h_{rA}$'s are possible, they are physically equivalent to $\xbar h_{rA} =0$. Second,  we explore an alternative suggested by the analysis of electromagnetism \cite{Henneaux:2018gfi} and of the $2$-form gauge field \cite{Henneaux:2018mgn}, in which the symplectic structure is modified by surface terms.  We find again that this option is physically equivalent to $\xbar h_{rA} = 0$.
\end{itemize} 

The non trivial structure at spatial infinity results from the fact that there are physical degrees of freedom - i.e., degrees of freedom invariant under ``small'' or ``proper'' gauge transformations - which are not invariant under ``large'' or ``improper'' gauge transformations.  The distinction between proper and improper gauge transformations is therefore crucial.  This distinction comes from the charges, which are defined through the standard Noether procedure from the symmetries of the action. For this to be the case, one needs a well defined (finite) action, which is indeed invariant.  The charges are identically zero (for the given set of boundary conditions) in the case of proper gauge transformations, or may be non zero in the case of improper gauge transformations.    The boundary conditions play an essential role, both in the definition of symmetries (which must leave the boundary conditions and the action invariant) and in the actual computation of the charges.

The impact of our boundary conditions on the structure of the physical states in the quantum theory, and in particular of the role of BMS charges in the dressings of physical operators \cite{Bousso:2017dny,Giddings:2019hjc},  is an important question which we leave for future investigation.

\section*{Acknowledgements}

MH acnowledges the hospitality of the Erwin Schr\"odinger International Institute for Mathematics and Physics while this paper was written. Work  partially supported by the ERC Advanced Grant ``High-Spin-Grav'' and by FNRS-Belgium (convention IISN 4.4503.15).

\begin{appendix}

\section{$2+1$ decomposition of the spatial geometry}
\label{app:2+1}

Polar coordinates are denoted in this appendix $x^i=(r,x^A)$
where $x^A$ are coordinates on the 2-sphere. We introduce  the ``lapse'' $\lambda$ and the ``shift'' $\lambda^A$ adapted to the slicing of space by the spheres of constant radius $r$, 
\begin{equation}
\gamma_{AB} \equiv g_{AB}, \quad \lambda_A \equiv g_{rA}, \quad \lambda \equiv \frac{1}{\sqrt{g^{rr}}}.
\end{equation}
In terms of these, he metric and its inverse take the form:
\begin{equation}
g_{ij}= \left(\begin{array}{cc}
\lambda^2 + \lambda_C\lambda^C & \lambda_B \\
\lambda_A & \gamma_{AB}
  \end{array}
\right),\quad 
g^{ij}= \left(\begin{array}{cc}
\frac{1}{\lambda^2} & -\frac{\lambda^B}{\lambda^2} \\
-\frac{\lambda^A}{\lambda^2} & \gamma^{AB}+ \frac{\lambda^A\lambda^B}{\lambda^2}
  \end{array}
\right),
\end{equation}
where we used $\gamma_{AB}$ and its inverse $\gamma^{AB}$ to raise and
lower the angular indices $A, B, ...$

Introducing the extrinsic curvature of the 2-spheres $K_{AB}$, we can write
all the Christoffel symbols:
\begin{eqnarray}
K_{AB} & =  & \frac{1}{2 \lambda} \left( - \d_r g_{AB} + D_A \lambda_B
  + D_B \lambda_A\right)\\
\Gamma^r_{AB} & = & \frac{1}{\lambda} K_{AB} \\
\Gamma^A_{BC} & = & {}^\gamma\Gamma^A_{BC} - \frac{\lambda^A}{\lambda} K_{BC} \\
\Gamma^r_{rA} & = & \frac{1}{\lambda} \left( \d_A \lambda + K_{AB} \lambda^B\right) \\
\Gamma^r_{rr} & = &\frac{1}{\lambda} \d_r \lambda
+\frac{\lambda^A}{\lambda} \left( \d_A \lambda + K_{AB}
  \lambda^B\right) \\ 
\Gamma^A_{rB} & = & -\frac{\lambda^A}{\lambda} \left( \d_B \lambda + K_{BC}
  \lambda^C\right) + D_B \lambda^A - \lambda K^A_B \\
\Gamma^A_{rr} & = & -\lambda \left( \gamma^{AB} + \frac{\lambda^A
    \lambda^B}{\lambda^2}\right) \left( \d_B \lambda +
  K_{BC}\lambda^C\right) - \lambda^C \left( D^A \lambda_C - \lambda
  K^A_C\right) \nonumber \\ && \qquad- \frac{\lambda^A}{\lambda} \d_r \lambda + \gamma^{AB}
\d_r \lambda_B
\end{eqnarray}
where $D_A$ is the covariant derivative associated to $\gamma_{AB}$.

The Ricci tensor is given by:
\begin{eqnarray}
	{}^{(3)}R_{AB} & = & \frac{1}{\lambda} \d_r K_{AB} + 2K_{AC}K^C_B -
	K K_{AB} - \frac{1}{\lambda} D_AD_B \lambda \nonumber \\ 
	&& \quad + {}^{\gamma} R_{AB} - \frac{1}{\lambda} 
	\cL_\lambda K_{AB},\\
	{}^{(3)}R_{rA} & = & \lambda \left( \d_A K - D_B
	K^B_A\right) + {}^{(3)}R_{AB} \lambda^B,\\
	{}^{(3)}R_{rr} & = & \lambda(\d_rK - \lambda^A\d_A K) - \lambda^2
	K^A_B K^B_A - \lambda D_AD^A\lambda \nonumber \\ && \quad-
	{}^{(3)}R_{AB}\lambda^A\lambda^B + 2 \, {}^{(3)}R_{rB}\lambda^B,
\end{eqnarray}
while the Ricci scalar takes the form
\begin{equation}
	{}^{(3)}R=\frac{2}{\lambda}(\d_rK -\lambda^A\d_AK) + {}^{\gamma}R-
	K^A_BK^B_A-K^2-\frac{2}{\lambda}D_AD^A\lambda.
\end{equation}

\section{Asymptotic transformation of the fields}
\label{app:AsymTransfFields}

The action of the asymptotic symmetries on the asymptotic fields can be computed from the variation of the metric and its momentum, and is given by
\begin{flalign}
	\delta_\xi \xbar k_{AB} & = \cL_Y \xbar k_{AB} + \xbar D_A \xbar D_B W + W
\xbar \gamma_{AB}\nonumber \\ &\qquad + \frac{b}{\sqrt{\xbar\gamma}} (\xbar \pi_{AB} - \xbar
\gamma_{AB} \xbar \pi^C_C) + \frac{1}{\sqrt{\xbar\gamma}} \xbar D_A( b \xbar
\pi^{rC} \xbar\gamma_{CB}) + \frac{1}{\sqrt{\xbar\gamma}} \xbar D_B( b \xbar
\pi^{rC} \xbar\gamma_{CA}),\\
	\delta_\xi \xbar\lambda &= \frac{b}{4\sqrt 
	{\xbar\gamma}} 
	\xbar p +  Y^C
	\d_C \xbar \lambda,\\
	\delta_\xi(\xbar \pi^{rr} - \xbar \pi^A_A) & = \cL_Y (\xbar \pi^{rr} -
	\xbar \pi^A_A)+\sqrt{ \xbar \gamma} \left( 2 b\xbar
	D_C \xbar D^C \xbar \lambda + 2 \xbar D^C b \d_C \xbar \lambda
+ 6 b \xbar \lambda\right) \\
	\delta_\xi \xbar \pi^{rA} & = \cL_Y \xbar\pi^{rA} + \sqrt{\xbar
	\gamma} \left(\xbar D_B(b \xbar k^{BA}) + \xbar D^A b \xbar k
	- \xbar D^A T\right), \\
	\delta_\xi \xbar \pi^{AB} & = \cL_Y \xbar \pi^{AB} +
	\sqrt{\xbar\gamma} \left( \xbar D^A \xbar D^B T - \xbar \gamma^{AB}
		\xbar D_C \xbar D^C T\right) + 3b \sqrt{\xbar\gamma}\left(\xbar k^{AB} - \xbar
	\gamma^{AB} \xbar k\right) \nonumber \\ & \qquad + \sqrt{\xbar\gamma} b
	\left(\xbar\gamma^{AB} \xbar D_C \xbar D^C \xbar k + \xbar D_C
		\xbar D^C \xbar k^{AB} - \xbar D_C \xbar D^A \xbar
	k^{CB} - \xbar D_C \xbar D^B \xbar k^{CA}\right)\nonumber\\ &
	\qquad +\sqrt{\xbar \gamma} \Big( - \xbar D^A b \xbar D^B
	\xbar k - \xbar D^B b \xbar D^A \xbar k+ \xbar \gamma^{AB} \xbar D_C
	b \xbar D^C \xbar k  + 2 \xbar \gamma^{AB} \xbar D^D \xbar k^C_D \d_C
	b\nonumber\\ & \qquad \qquad - \xbar D^A \xbar k^{BC} \d_Cb-
\xbar D^B \xbar k^{AC} \d_Cb + \xbar D^C \xbar k^{AB} \d_C b\Big).
\end{flalign}

One can read from these transformation rules that the variation of the functions $(U)^{odd}$ and $(V)^{even}$ take the following form:
\begin{flalign}
	\delta_\xi (U)^{odd} & = Y^C \d_C (U)^{odd} - b (V)^{even} 
			+ (W)^{odd},\\
	\delta_\xi (V)^{even} & = Y^C \d_C (V)^{even} - 3 b
			(U)^{odd} - \d_A b \xbar D^A (U)^{odd} - b \xbar D_A
			\xbar D^A (U)^{odd} + (T)^{even}.
\end{flalign}
These transformations have an inhomogeneous component ($(W)^{odd}$ for $U$ and $(T)^{even}$ for $V$).  It follows that if one starts from a configuration that satisfies the untwisted parity conditions, one generically generates a nonvanishing twist that takes exactly the prescribed form, except if one restricts the transformation to the Poincar\'e algebra in which case the twist remains zero.  Invariance of the boundary conditions under the extended set of transformations is in that sense direct.  

\section{More on the integrability conditions of the boost charges}
\label{sec:IntegrabilityLambda}

\subsection{Further study of the integrability of (\ref{eq:KFull})}

We further analyse here the conditions implied by the requirement that (\ref{eq:KFull}) should be integrable.  We argue, in the symplectic framework adopted in the paper, that the variables $\xbar \lambda^A$ actually do not need to vanish, but that one can set them equal to zero by a coordinate redefinition that does not affect the charges (proper gauge transformation).   

For the last line in (\ref{eq:KFull}) to be integrable, one must have
\be
\oint d^2x \, 	
	  \sqrt{\xbar\gamma}\,  \Big(d_V\xbar \lambda^C\, \d_C
	b \,\xbar\gamma^{AB} - b\xbar D^A (d_V \xbar\lambda^B) \Big) \, d_V   \xbar h_{AB} = 0. \label{eq:KFullPart}
\ee
The even part of $\xbar h_{AB}$ is arbitrary, which forces the even part of $\xbar\lambda^A$ to vanish\footnote{More precisely, $d_V (\xbar \lambda^A)^{even} = 0$, which implies that $(\xbar \lambda^A)^{even}$ should be field-independent.  There is no field-independent non-vanishing vector field to which $(\xbar \lambda^A)^{even}$ could be equated without conflicting with the symmetries, and so we take $(\xbar \lambda^A)^{even}= 0$. }.The odd part of $\xbar h_{AB}$ must have the form of equation (\ref{eq:pParitySpher002}) from which it follows upon integration by parts and use of $\xbar D_A \xbar D_B b + \xbar \gamma_{AB} b = 0$, that 
\be
\oint d^2x \, 	
	  \sqrt{\xbar\gamma}\,  \Big(  - \xbar D_A b\,  \xbar D^A d_V \Lambda - b \, \xbar \Delta d_V \Lambda - 4 b \, d_V \Lambda  \Big) \, d_V   \xbar U = 0. \label{eq:KFullPartOdd}
\ee
with
\be
\Lambda \equiv \xbar D_A (\xbar \lambda^A)^{odd}.
\ee
For (\ref{eq:KFullPartOdd}) to hold with unrestricted (odd) $\xbar U$'s, one must have
\be 
- \xbar D_A b\,  \xbar D^A d_V \Lambda - b \, \xbar \Delta d_V \Lambda - 4 b \, d_V \Lambda = 0.
\ee
Taking for $b$ a generic boost, one easily concludes that $d_V \Lambda$ should vanish.  This implies that $\Lambda$ should be field-independent.  The simplest choice is $\Lambda = 0$, which yields
\be
\xbar \lambda^A =(\xbar \lambda^A)^{odd} , \; \; \; \xbar D_A (\xbar \lambda^A)^{odd} = 0. \label{eq:KFullPartOddZero}
\ee
Other choices with $\xbar D_A (\xbar \lambda^A)^{odd} \not=0$ could be envisioned, but we have not investigated their consistency.

We now explore  the consequences of (\ref{eq:KFullPartOddZero}).  When this condition holds, the last line in (\ref{eq:KFull}) is not only integrable, but in fact zero as can be seen by making the same integration by parts.  The variables $\xbar \lambda^A$ do not contribute to the charges. One can actually set them equal to zero by a proper gauge transformation.  Indeed, if we redefine the angles as
\be
x'^A = x^A + \frac{1}{r} \xbar \zeta^A, \; \; \; \xbar \zeta^A = \hbox{odd},  \; \; \; D_A \xbar \zeta^A = 0
\ee
we get
\be
\xbar \lambda^A  \rightarrow \xbar \lambda^A + \xbar \zeta^A
\ee
and so the form (\ref{eq:KFullPartOddZero}) of $\xbar \lambda^A$ is preserved.  Taking $\xbar \zeta^A = - \xbar \lambda^A $ sets the new $\xbar \lambda^A$ equal to zero.  This transformation is permissible, because its charge
\be
\oint \xbar \zeta^A \xbar \gamma_{AB} \xbar \pi^{rB}
\ee
vanishes.  Indeed, given that $\xbar \zeta^A$ is odd, only the odd part of $\xbar \pi^{rB}$ contributes to the integral, which becomes
\be
\oint d^2 x \, \xbar \zeta^A \, \xbar \gamma_{AB} \, \xbar \pi^{rB} = - \oint d^2 \, x\sqrt{\xbar \gamma}\, \xbar \zeta^A \, \xbar D_A V = \oint d^2x \, \sqrt{\xbar \gamma}\,  \xbar D_A \xbar \zeta^A  \, V = 0.
\ee

\subsection{Modifying the symplectic structure by a surface term}

There is  another option for guaranteeing the existence of well defined charges for the 
boosts, which consists in modifying the symplectic structure by a surface term. In the analysis of the asymptotic 
symmetries of electromagnetism \cite{Henneaux:2018gfi} and of the $2$-form gauge fields \cite{Henneaux:2018mgn}, we encountered similar 
integrability problems for the boosts charges. In these cases, where extra restrictions on the asymptotic fields would 
remove physically important symmetries, the solution proposed was to add a boundary term to the symplectic 
structure. A similar approach is also available here but, as we will 
see below, it is again equivalent to the solution obtained by imposing $\xbar\lambda^A=0$.

The asymptotic conditions considered in this subsection are the
generalized parity conditions given in
\eqref{eq:asymptgrav0a}-\eqref{eq:asymptgrav0c} and
\eqref{eq:asymptgrav0d}-\eqref{eq:asymptgrav0f} together with the
asymptotic constraints given in \eqref{eq:asymptgrav1}. But we do {\it not} impose (\ref{eq:asymptgrav2a})-(\ref{eq:asymptgrav2b}). 

We first provide the form of the alternative solution, verify then that the boost charges are well-defined, and comment finally on the physical equivalence with the simpler solution $\xbar \lambda^A = 0$.

The action is modified by the surface term 
$$
2 \oint d^2x \, \xbar \pi^{rA} \xbar \gamma_{AB} \d_t \xbar \lambda^B
$$
involving the time derivatives of $\xbar \lambda^A$.
It thus takes the following form (including the energy term, since the lapse goes to one at infinity)
\begin{multline}
	S[g_{ij}, \pi^{ij}, N, N^i]  =\int dt \left\{ \int d^3x \left( \pi^{ij} \d_t g_{ij}  - N^i \mathcal
		H^{grav}_i - N \mathcal H^{grav} \right) \right. \\ \left.
		+ 2 \oint d^2x \left(\xbar \pi^{rA} \xbar \gamma_{AB} \d_t \xbar \lambda^B -\sqrt{\xbar \gamma}\, \xbar h_{rr} \right) \right\},
\end{multline}
\begin{gather}
	\mathcal H^{grav}  = - \sqrt g R + \frac{1}{\sqrt g} (\pi^{ij} \pi_{ij} -
	\frac{1}{2} \pi^2),\quad
	\mathcal H^{grav}_i  = -2 \nabla_j \pi^j_i, \\
	N = 1 + O(r^{-1}), \quad N^r = O(r^{-1}), \quad N^A = O(r^{-2}).
\end{gather}
The variation of the surface  contribution to the kinetic term
 produces two apparently extra equations of motion
\begin{equation}
    \d_t \xbar \pi^{rA} = 0, \qquad \d_t \xbar \lambda^A = 0.
\end{equation}
However, as these two equations are already implied by the bulk equations of motion, the dynamics are unchanged. As a result of the modification of the action,   the symplectic structure picks up a boundary term:
\begin{equation}
    \Omega = \int d^3x \, d_V \pi^{ij} d_V g_{ij} 
      + 2 \oint d^2x \, d_V \xbar \pi^{rA} \xbar \gamma_{AB} d_V  \xbar \lambda^B
\end{equation}
and does not take the standard canonical ``Darboux'' form.

The boundary conditions are preserved by vector fields behaving asymptotically as 
\begin{gather}
    \xi = rb + F_{(b)} + O(r^{-1}), \quad \xi^r = W + O(r^{-1}), \quad \xi^A = Y^A + r^{-1} I^A + O(r^{-2}), (\label{app:AsympSym0})\\
    F_{(b)}  = T- b \xbar k - \xbar \lambda^C \d_C b , \qquad
    \xbar D_A \xbar D_B b + \xbar \gamma_{AB} b = 0, \qquad \xbar D_A Y_B + \xbar D_B  Y_A= 0.
    \label{app:AsympSym1}
\end{gather}
Compared with the BMS4 generators considered in the main text, we see that the subleading term in $\xi$ takes an
extra contribution dependent on $\xbar \lambda^A$ while the subleading term in $\xi^A$ is now completely arbitrary. This is because the condition $\xbar h_{rA} = 0$ is not imposed, so that the condition $W_i = \partial_i (rW)$ of (\ref{eq:As3}) does not arise.  The corresponding transformations on the asymptotic fields $\xbar \lambda^A$ and $\xbar \pi^{rA}$ are given by
\begin{flalign}
\label{eq:varIlambda}
	\delta_\xi \xbar \lambda^A & = \frac{2b}{\sqrt {\xbar\gamma}} \xbar \pi^{rA} + \mathcal L_Y \xbar \lambda^A
	   + \xbar D^A W - I^A,\\
	   \label{eq:varIpi}
	\delta_\xi \xbar \pi^{rA} & = \cL_Y \xbar\pi^{rA} + \sqrt{\xbar \gamma} \left(\xbar D_B(b \xbar k^{BA}) 
	   + \xbar D^Ab \xbar k\right) 
	   - \sqrt{\xbar \gamma}\, \xbar D^A \left(F_{(b)} +b \xbar k+ b \xbar \lambda + \xbar \lambda^B \d_B b\right).
\end{flalign}
The action of the symmetries on the other asymptotic fields will present similar modifications but we will not 
need their explicit expressions to compute the charges. 

The presence of a boundary term in the symplectic structure implies that the standard 
prescription of \cite{Regge:1974zd} to define hamiltonian generators must be amended to take into account the non-Darboux form of $\Omega$. The simplest way to work out the modification is to go back to the 
original definition of a hamiltonian generator: a well defined generator $G_Q$ is associated to the
transformation $\delta_Q$ iff it satisfies the identity $-i_Q \Omega = d_V G_Q$. Using the result obtained
in section \ref{sec:sphericalcoords} and the variations \eqref{eq:varIlambda} and \eqref{eq:varIpi}, one
can then show that the asymptotic symmetries given above, and in particular the boosts, are hamiltonian. Their generators are given by
\begin{equation}
    G_{\xi} = \int d^3x \left( \xi \mathcal H + \xi^i \mathcal H_i\right) + \mathcal B_\xi,
\end{equation}
where the boundary term takes the form
\begin{multline}
    \mathcal B_\xi = \oint d^2x \Big\{2 Y^A (\xbar h_{AB} \xbar \pi^{rB} + \xbar \gamma_{AB}
	\pi^{(2)rB} + \xbar \lambda_A \xbar \pi^{rr})+ 2 W  \xbar \pi^{rr} + 2 \d_A W \xbar \pi^{rA}+ 2\mathcal L_Y \xbar \lambda^A\xbar\gamma_{AB}\xbar \pi^{rB}\Big\}\\
	- \oint d^2x \Big\{ 
	  -\frac{2b}{\sqrt {\xbar\gamma}} \xbar \pi^{rA} \xbar\gamma_{AB} \xbar \pi^{rB} 
	  - 2 \sqrt{\xbar\gamma}b k^{(2)} - 4\sqrt{\xbar\gamma} T \xbar \lambda
	  - \frac b 4 \sqrt{\xbar\gamma} (\xbar h^2 + \xbar h^A_B \xbar h^B_A) \\
	  + 6\sqrt{\xbar \gamma} b\, \xbar\lambda \, \xbar D_B \xbar \lambda^B
	   + 2\sqrt{\xbar \gamma} \xbar
	\lambda^D\d_D b \xbar k-2\sqrt{\xbar \gamma} 
	b\, \xbar D^A \xbar \lambda^B \xbar k_{AB} \\
	  + \sqrt{\xbar \gamma}\, b (\xbar D_A \xbar \lambda^A)^2 + \sqrt{\xbar \gamma} b\, \xbar D^A \xbar \lambda^B \xbar D_{(A} \xbar\lambda_{B)}
	  \Big\}.
\end{multline}
The full group of asymptotic symmetries (\ref{app:AsympSym0})-(\ref{app:AsympSym1}) is thus canonically realized.

The existence of these generators does not mean, however, that the asymptotic symmetry algebra is now bigger than BMS4. 
In fact, one can see that the parameter $I^A$ does not contribute to the
boundary term, which means that the associated transformation is a proper
gauge transformation. If we use this transformation to gauge fix
$\xbar\lambda^A=0$, the action and the generators of the BMS4 algebra
reduce to the ones we considered in the main text. As we announced earlier, the
two solutions to the integrability problem of the boost charges presented
in this work, namely imposing $\xbar\lambda^A=0$ or introducing a boundary term in the symplectic structure, are physically equivalent.  This gives additional support to the assumption $\xbar \lambda^A = 0$ made in the text.

\end{appendix}

\end{document}